\documentclass[12pt]{article}
\usepackage{hyperref}
\usepackage{cite}
\usepackage{xcolor}
\usepackage{graphicx}
\usepackage{caption,subcaption}
\usepackage{amsmath}
\usepackage{amssymb}
\usepackage{xspace}
\usepackage{verbatim}
\usepackage{mathtools}
\usepackage{tikz-cd}
\usepackage{braket}
\usepackage[normalem]{ulem}

\makeatletter
\@addtoreset{equation}{section}

\makeatletter
\renewcommand\section{\@startsection {section}{1}{\z@}%
                                   {-3.5ex \@plus -1ex \@minus -.2ex}
                                   {2.3ex \@plus.2ex}%
                                   {\normalfont\large\bfseries}}
\renewcommand\subsection{\@startsection{subsection}{2}{\z@}%
                                     {-3.25ex\@plus -1ex \@minus -.2ex}%
                                     {1.5ex \@plus .2ex}%
                                     {\normalfont\bfseries}}

\def\baselinestretch{1.2}
\newcommand{\be}{\begin{equation}}
\newcommand{\ee}{\end{equation}}
\newcommand{\beq}{\begin{eqnarray}}
\newcommand{\eeq}{\end{eqnarray}}

\newcommand{\gone}[1]{{}}

\definecolor{amber}{rgb}{1.0, 0.75, 0.0}

\begin{document}
\begin{titlepage}
\begin{flushright}
MAD-TH-17-11
\end{flushright}

\vfil

\begin{center}

{\bf \Large
Persistent Homology and Non-Gaussianity
}

\vfil

Alex Cole and Gary Shiu

\vfil

Department of Physics, University of Wisconsin, Madison, WI 53706, USA

\vfil
\end{center}

\begin{abstract}

\noindent In this paper, we introduce the topological persistence diagram as a statistic for Cosmic Microwave Background (CMB) temperature anisotropy maps. A central concept in `Topological Data Analysis' (TDA), the idea of persistence is to represent a data set by a family of topological spaces. One then examines how long topological features `persist' as the family of spaces is traversed. We compute persistence diagrams for simulated CMB temperature anisotropy maps featuring various levels of primordial non-Gaussianity of local type. Postponing the analysis of observational effects, we show that persistence diagrams are more sensitive to local non-Gaussianity than previous topological statistics including the genus and Betti number curves, and can constrain $\Delta f_{NL}^{\rm loc}= 35.8$ at the 68\% confidence level on the simulation set, compared to $\Delta f_{NL}^{\rm loc}= 60.6$ for the Betti number curves. Given the resolution of our simulations, we expect applying persistence diagrams to observational data will give constraints competitive with those of the Minkowski Functionals. This is the first in a series of papers where we plan to apply TDA to different shapes of non-Gaussianity in the CMB and Large Scale Structure.
\end{abstract}
\vspace{0.5in}

\end{titlepage}
\renewcommand{\baselinestretch}{1.05}  
\begin{section}{Introduction}
By postulating a period of accelerated expansion in the early universe,
inflation famously solves the flatness, homogeneity, and monopole problems of big bang cosmology
\cite{PhysRevD.23.347,Starobinsky:1980te,Linde:1981mu,Albrecht:1982wi}. 
But what establishes its role in modern precision cosmology is its prediction of an almost scale-invariant primordial density perturbation spectrum \cite{mukhanov1981quantum,mukhanov1982vacuum,hawking1982development,starobinsky1982dynamics,guth1982fluctuations,bardeen1983spontaneous,mukhanov1985gravitational}.
 These perturbations are not perfectly Gaussian, but the deviation from Gaussianity is typically small for single field slow-roll inflation \cite{Maldacena:2002vr}. 
 Nonetheless, it was shown that a broader class of inflationary models, even within the single field context, can produce significant levels of non-Gaussianity with distinctive `shapes' (functional dependences on momenta)  \cite{Chen:2006nt}.
 By now, many inflationary mechanisms generating a wide variety of non-Gaussian shapes are known (see \cite{Chen:2010xka,Renaux-Petel:2015bja} for reviews).
 Since non-Gaussianity contains a functional worth of information, it can discriminate sharply between models.
 Thus experimental constraints on primordial non-Gaussianity can narrow down the robust swath of models possibily describing the inflation of our universe.

With advances in experimental cosmology,
we 
now
have access to the statistics of the primordial density perturbations via Cosmic Microwave Background (CMB) measurements. The perturbations imprint themselves onto the CMB in the form of temperature and polarization anistropies. Thus we may reconstruct features of the primordial density perturbations by studying the statistics of the CMB. Measuring primordial non-Gaussianity in the CMB has proven to be difficult, however, with the non-Gaussian component of CMB data dominated by observational effects \cite{Ade:2015ava}. Moreover, while a perfectly Gaussian probability distribution function can be described by just its average and standard deviation, there are infinitely many ways for a random variable to deviate from Gaussianity. In other words, there is no single signature of primordial non-Gaussianity for which we may simply test. We must instead look for non-Gaussianity `from all sides', employing complementary approaches.

This complementarity of statistical measures currently features two prongs.  In harmonic space approaches, one studies the shapes of three-point and higher-order correlation functions in momentum space, fitting against templates for different known models \cite{Babich:2004gb,Liguori:2005rj}. The other approach involves studying the geometric and topological features of excursion sets of temperature and polarization anisotropies of the CMB. One class of geometric observables is the Minkowski Functionals (MFs) \cite{Mecke:1994ax,1538-4357-482-1-L1,1998MNRAS.297..355S,WINITZKI199875}. Work has been done to give analytic expressions for the MFs of weakly non-Gaussian fields \cite{Hikage:2006fe}. One of the MFs, known as the genus \cite{1986ApJ...306..341G}, is topological in nature. Up to a numerical factor, it is the Euler characteristic, which can be calculated as the alternating sum of Betti numbers. Thus in \cite{2012ApJ...755..122C} it was proposed to use the Betti number curves themselves rather than the genus curve for CMB studies. The authors explicitly demonstrated using CMB simulations that the Betti number curves contain more information about non-Gaussianity than the genus curve.

In this paper, we propose a strengthening of the topological approach using techniques from `Topological Data Analysis' (TDA) \cite{carlsson2009topology,zomorodian2005topology,edelsbrunner2010computational}. Specifically, we invoke persistent homology to further strengthen the statistical power of the Betti number curves. Given a data set, the idea of persistent homology is to consider a family of topological spaces, called a filtration. Each topological space in the filtration represents the data set. In each of these spaces one may compute the homology group, which, roughly speaking, counts $n$-dimensional `holes' in the space. However, the homology of one particular representation in the filtration does not give us very much information about the data set, as it depends on the specific way we choose to represent the data. The power of persistence is that one tracks the lifetime of individual homological features as one moves through the filtration of topological spaces. Thus persistence accesses a layer of information that is invisible to the Betti number curves, which merely count the total number of distinct homological features at a given step in the filtration. In the language of persistent homology, we track the `births' and `deaths' of these topological features. The scatter plot of births and deaths for a filtration is called a persistence diagram. While one may calculate the Betti number curves from a persistence diagram, the reverse is not true. Thus peristence diagrams contain strictly more information than the Betti number curves. We therefore expect the addition of persistence to sharpen our topological studies of the CMB.

In this paper, we review the concept of persistent homology and describe its application to CMB temperature anisotropy data. We compute the persistence diagrams of sublevel filtrations from a publicly available set of simulations \cite{0067-0049-184-2-264} describing primordial non-Gaussianity of local type \cite{Komatsu:2001rj}. We then explicitly demonstrate using a likelihood function analysis that persistence diagrams contain more information about local primordial non-Gaussianity than the Betti number curves. Specifically, we are able to constrain $\Delta f_{NL}=35.8$ at 68\% confidence on our simulation set, almost a factor of two better than the Betti number curves. In other words, persistence provides a promising strengthening of topological techniques in the search for non-Gaussianity.

The present work is the first in a series of papers in which we apply TDA to cosmological datasets. As such, we consider the simplest setup in order to illustrate our approach and to compare with other methods used in existing literature. As  a proof of concept, we use simulations for only local-type  primordial non-Gaussianity (which are readily available) and defer observational effects for future work.
Natural next steps are to apply our method to study other shapes (especially those hard to detect via fitting templates of the bispectrum), to include polarization degrees of freedom, and to use actual experimental data. 
We also plan 
to apply persistent homology to large-scale structure data. Here the topology is richer (and the dimension of the space higher), with interlocking filaments, walls, and voids making up what is called the `cosmic web' \cite{Pranav:2016gwr}. Our findings for these extensions of the present work will be reported elsewhere.

This paper is organized as follows. In Section 2 we give a quick introduction to persistent homology and persistence diagrams. In Section 3 we review local non-Gaussianity and the simulations we use. In Section 4 we describe our numerical pipeline and detail our results. We end this paper with conclusions and outlook in Section 5.
\end{section}
\begin{section}{Persistent homology}
In this section we describe persistent homology and how it can be applied to CMB temperature anisotropy data. First we briefly review how homology formalizes the notion of `counting holes' in a topological space in the context of simplicial homology. We then describe how the extra ingredient of persistence allows one to study the topology of discrete data sets in a suitably stable way. Persistence diagrams arise as a natural way to represent the outcome of persistent homology calculations. It is these persistence diagrams we will use as statistics for our data. We then outline how persistent homology can be used to study the CMB and constrain cosmological parameters.

\begin{subsection}{Simplicial homology}
Homology is a standard technique for identifying a topological space (see e.g. \cite{zomorodian2005topology}). Roughly speaking, this is done by counting `holes' of various dimension in the space. In practice this amounts to identifying special `loops' in the space. Specifically, one looks for `loops' that may not be continuously shrunk to a point (these are the loops wrapping holes) -- each `independent' one of these contributes to the homology of the space. Here we will formalize these notions in the context of simplicial homology. As continuous spaces are often unwieldy for calculations, one often resorts to discrete representations. Moreover, for the purpose of data analysis, discrete representations are natural. We will represent our data using simplicial complexes and perform our topological calculations in the context of simplicial homology.

Simplicial complexes are made up of simplices. Low-dimensional simplices include
\begin{itemize}
	\item vertices, or 0-simplices;
	\item edges, or 1-simplices;
	\item triangles, or 2-simplices;
	\item tetrahedra, or 3-simplices,
\end{itemize}
and so on. Often we will work in arbitrary dimension, where we will refer to $k$-simplices. Each $k$-simplex contains lower-dimensional simplices on its boundary. For example, a triangle contains three edges and three vertices. We call these lower-dimensional simplices \emph{faces} of the higher-dimension simplex.\footnote{A simplex also has itself as a face. We say $\sigma$ is a \emph{proper} face of $\tau$ if $\sigma\neq \tau$.}  A simplicial complex $S$ is a set of simplices that is closed under intersection of simplices and closed under taking faces of simplices. In other words,
\begin{align}
	\sigma,\tau\in S\implies \sigma\cap\tau\in S,\\
	\tau\subseteq\sigma,\sigma\in S\implies \tau\in S
\end{align}
Examples of simplicial complexes and collections of simplices that are not simplicial complexes are shown in Figure \ref{fig:simp}.
\begin{figure}\centering
	\includegraphics[width=0.4\textwidth]{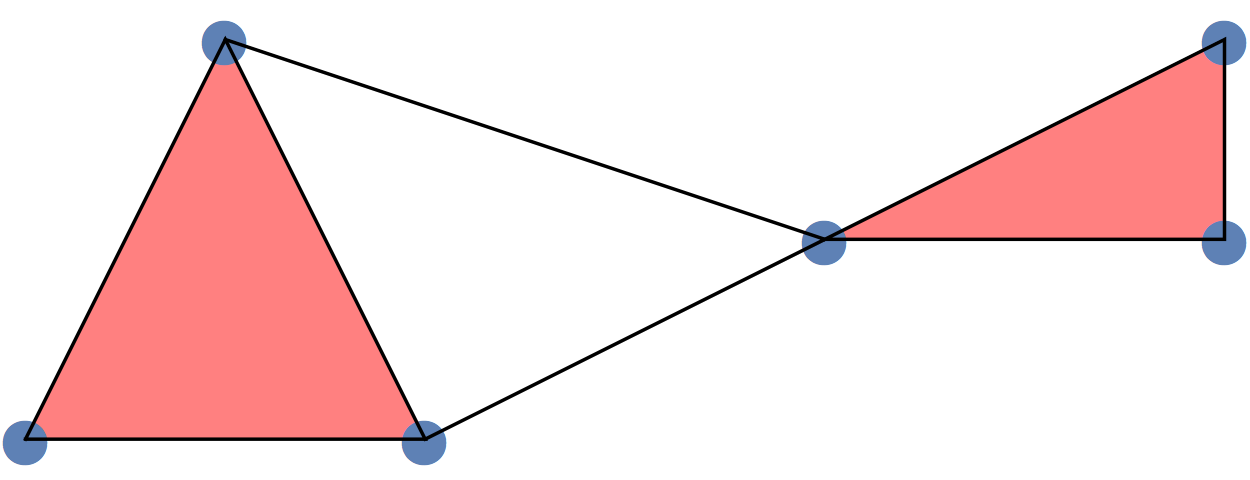}\qquad\qquad\includegraphics[width=0.4\textwidth]{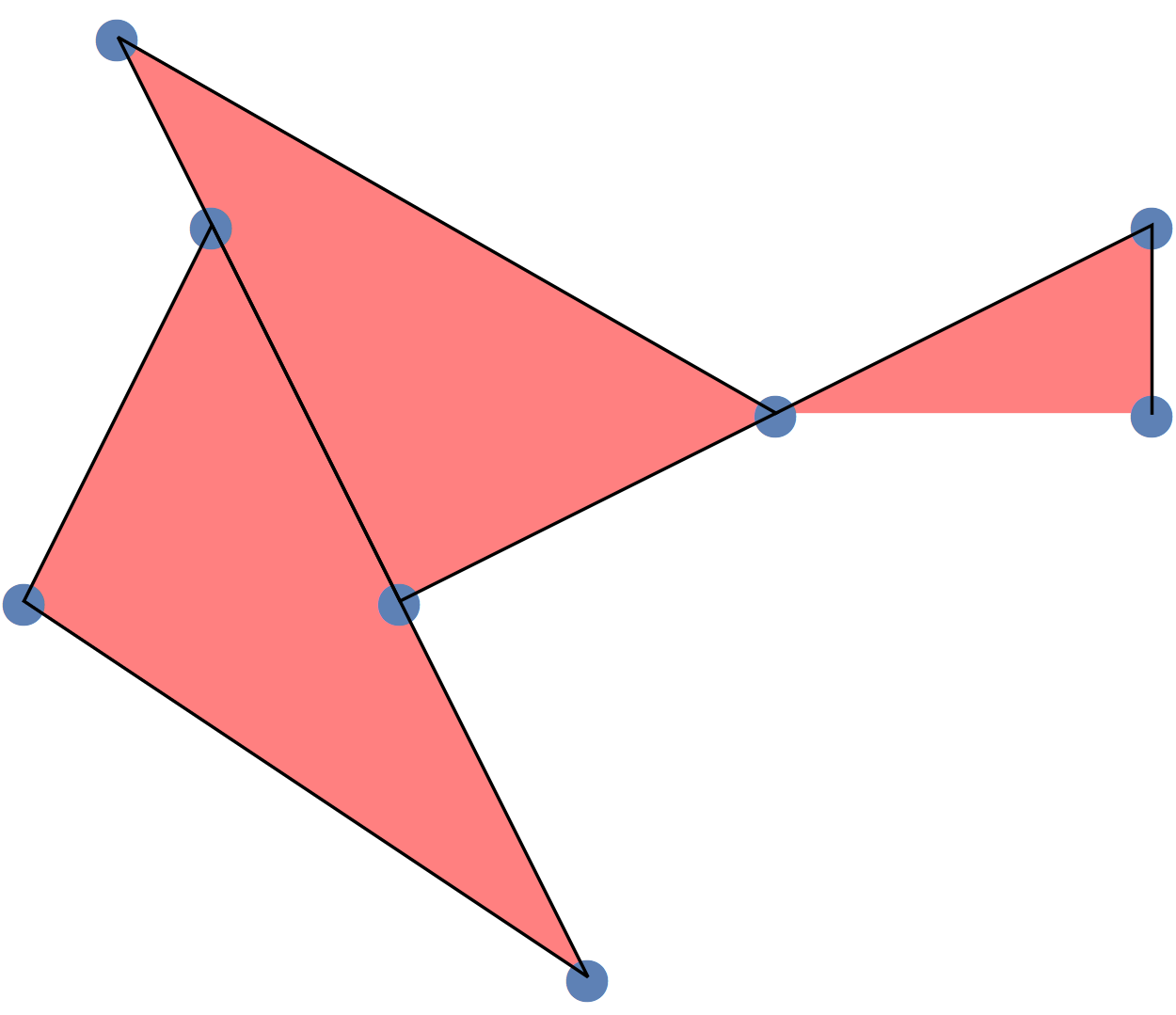}\caption{On the left is a simplicial complex. It is closed under taking intersections and faces of simplices. On the right, the intersection of the two large triangles is not a simplex and the one of the small triangle's edges is not present, so the collection is not a simplicial complex.}\label{fig:simp}
\end{figure}
Given a simplicial complex $S$, we define $k$-chains as collections of $k$-simplices in $S$. Each of these may be formally represented as a sum
\begin{align}
	\sum_{i}a_i\sigma_i,\quad a_i\in\mathbb{Z}_2
\end{align}
where $i$ runs over the set of $k$-simplices in $S$ and we use $\mathbb{Z}_2$-valued coefficients.\footnote{One may more generally use a field other than $\mathbb{Z}_2$. Here we stick with $\mathbb{Z}_2$, which is generally the most practical computational choice.} Under element-wise addition, the $k$-chains form a group, which we denote as $C_k$. 

The group of $k$-chains has two important subgroups. To define these, we must first define the \emph{boundary map}. The boundary map takes a $k$-chain to a $(k-1)$-chain in the following way. Write a $k$-simplex in terms of its vertex set $\sigma=[v_0,v_1,\dots, v_k]$. Then the boundary of $\sigma$ is
\begin{align}
	\partial_k\sigma=\sum_{j=0}^k[v_0,\dots,\hat{v}_j,\dots,v_k]
\end{align}
where the hatted vertex is omitted. The boundary operator is then defined on $k$-chains by linear extension. The boundary map $\partial_k$ is a homomorphism from $C_k$ to $C_{k-1}$. 

A $k$-cycle is a $k$-chain with empty boundary, $\partial_k \sigma=0$. By the linearity of $\partial_k$, the $k$-cycles form a group, denoted $Z_k$. Put succinctly, $Z_k=\ker \partial_k$. Analogously we have the group of $k$-boundaries, denoted $B_k$, defined as the image of the $(k+1)$-th boundary map, $B_k=\textrm{im}~\partial_{k+1}$. In other words, for each $k$-boundary $\sigma$ there exists a $(k+1)$-chain $\tau$ such that $\partial_{k+1}\tau=\sigma$. 

It turns out that $B_p\subseteq Z_p$. This result is called the Fundamental Lemma of Homology. In words, it is the statement that the boundary of a boundary is empty. In other words, every $k$-boundary is also a $k$-cycle. However, the reverse is not true. In fact, cycles that are not boundaries wrap `holes' in our complex. Thus we define the $k$-th homology group as the the $k$-cycle group modulo the $k$-boundary group, $H_k=Z_k/B_k$. The $k$-th Betti number is this group's rank, $\beta_k=\textrm{rank}~H_k$. One can view the elements of $H_k$ as $k$-cycles under the equivalence relation $\sigma\sim \tau$ if $\sigma=\tau+b$ for some $k$-boundary $b$. In this way we can regard homology as counting `independent' features. Intuitively, $\beta_0$ counts the connected components of a simplicial complex, while $\beta_k$ for $k>0$ counts $(k+1)$-dimensional holes.

Data sets often display some topological structure. We would like to characterize the topology of a data set using homology. Doing so robustly, as explained in the next section, requires a new ingredient, persistence.

\end{subsection}
\begin{subsection}{Persistent homology}
Consider some point cloud $D\subset \mathbb{R}^2$, like the one depicted in Figure \ref{fig:ann}. We might choose to study the point cloud via topological means. To do so, we can form a simplicial complex corresponding to the point cloud and compute its homology. However, this procedure is far from unique, as there is significant ambiguity in how a point cloud `corresponds' to a simplicial complex. For example, it is natural to represent each point via a vertex, but we must start making more difficult choices when it comes to connecting the vertices with edges. These choices can result in different topological invariants for simplicial complexes representing the same data set, an undesirable outcome. This procedure is also not necessarily stable against perturbations of the data set. We would like our computation to avoid representational ambiguities and be stable against perturbations to the data.

\begin{figure}
\centering
\includegraphics[width=0.5\textwidth]{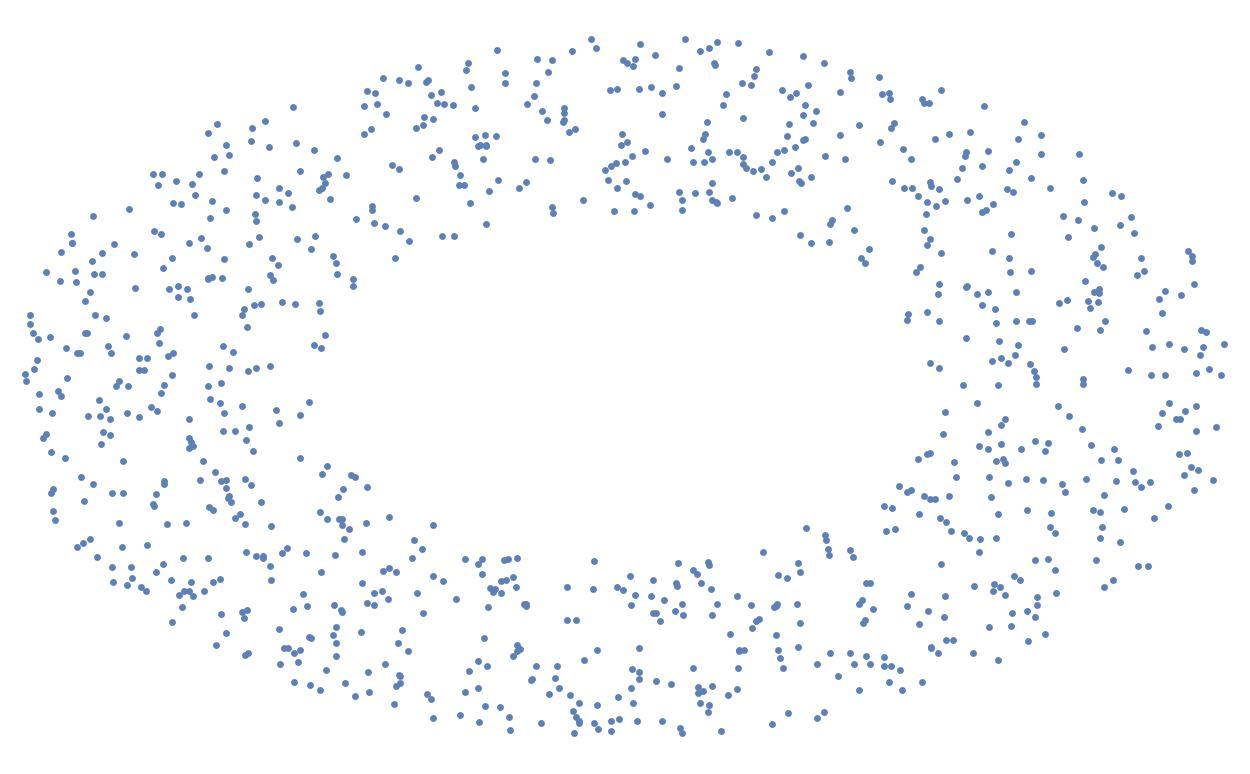}
\caption{Points randomly sampled from an annulus. The data set has a clear topological character, with $\beta_0=1$ and $\beta_1=1$. Persistent homology can formally describe the data's topology by using a sequence of simplicial complexes.}\label{fig:ann}
\end{figure}

A procedure that achieves these goals is \emph{topological persistence}. The idea of topological persistence is to consider a nested sequence of topological spaces
\begin{equation}
	\mathbb{M}_0\subseteq \mathbb{M}_1\subseteq\dots\subseteq \mathbb{M}_n
\end{equation}
and track topological features calculated via their homology groups
\begin{equation}\label{eqn:pHom}
	H_k(\mathbb{M}_0)\to H_k(\mathbb{M}_1)\to \dots \to H_k(\mathbb{M}_n)
\end{equation}
The sequence of topological spaces is called a \emph{filtration}. As we move through the filtration, different topological features appear (e.g. connected components are added to the space, 1-cycles form), while other features disappear (e.g. 1-cycles are `filled in' and are no longer nontrivial in the homology). We refer to these instances as \emph{births} and \emph{deaths}, respectively. When two topological features merge (e.g. two formerly disconnected components are joined), we adopt the \emph{elder rule}: the feature that was born earlier remains while the younger feature dies. Tracking the births and deaths of topological features allows us to resolve the strong features of a perhaps noisy topological space. As this calculation involves tracking how long homological features persist, it is referred to as persistent homology\footnote{One may also consider persistent cohomology, or other generalizations to the sequence in (\ref{eqn:pHom}) like zigzag persistence \cite{2008arXiv0812.0197C}.}. This procedure partially resolves the problem of representational ambiguity by considering a family of topological spaces instead of a single representation. Persistent homology (represented via persistence diagrams, see Section \ref{PDs}, and with a suitable metric) has also been proven stable against perturbations to the data set \cite{cohen2005stability}. 

Moving through the filtration often corresponds to scanning over lengths. In this sense persistence homology is a multi-scale approach, capable of identifying structures of various scales. While the annular character of Fig. \ref{fig:ann} is clear, there may be other topological structures living at smaller scales. These smaller-scale structures often contain important information about the data set.

In this paper we focus on the \emph{sublevel filtration}. Consider a smooth $d$-manifold $\mathbb{M}$ and a smooth function $f:\mathbb{M}\to\mathbb{R}$ defined on the manifold. The sublevel set corresponding to filtration parameter $\nu$ is
\begin{equation}
	f^{-1}(-\infty,\nu]=\{x\in\mathbb{M}~|~f(x)\leq \nu\}
\end{equation}
These are sometimes referred to as excursion sets, although we shall use the term `sublevel' to distinguish the direction in which they are bounded. (One can analogously define superlevel sets.) Since $f^{-1}(-\infty,\nu_1]\subseteq f^{-1}(-\infty,\nu_2]$ whenever $\nu_1\leq \nu_2$, the sequence of sublevel sets for a function $f$ on $\mathbb{M}$ defines a filtration. The sublevel filtration for a function defined on an interval of $\mathbb{R}$ is shown in Figure \ref{fig:sub}.
\begin{figure}[h]\centering
	\includegraphics[width=0.45\textwidth]{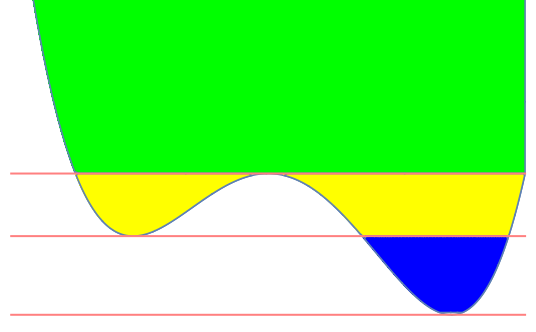}\caption{Illustration of the sublevel filtration for a 1-dimensional function. The homology of the sublevel sets changes at the function's critical points. When our filtration parameter (represented by a horizontal line) is in the blue region, the sublevel set has one connected component, ${\color{blue}\beta_0}=1$. In this region, we `know' about the lowest minimum. In the yellow region, the sublevel set has two connected components, ${\color{amber}\beta_0}=2$. When we reach the green region, the two minima merge, and we have once again have one connected component, ${\color{green}\beta_0}=1$.}\label{fig:sub}
\end{figure}
 In this paper, our manifold $\mathbb{M}\subset\mathbb{R}^2$ is a rectangular patch of a 2-sphere and $f$ is the temperature anisotropy of the CMB at that point on the sky. We work in rectangular patches rather than on the entire 2-sphere since CMB data is generally incomplete, for example in the galactic plane. An example of the sublevel filtration for such a data set is shown in Figure \ref{fig:sub2d}.

 \begin{figure}[h]
 \includegraphics[width=0.25\textwidth]{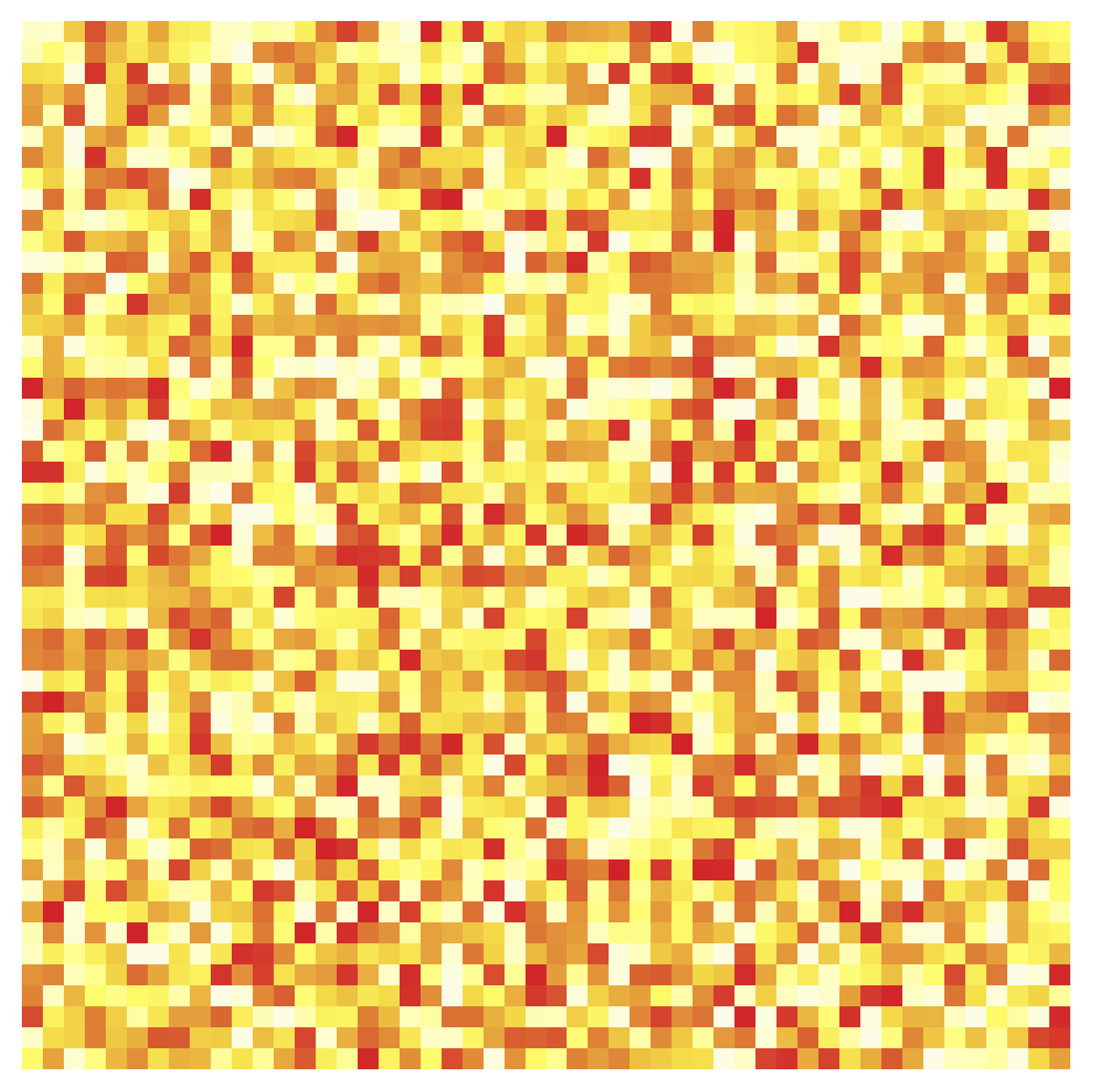}\includegraphics[width=0.25\textwidth]{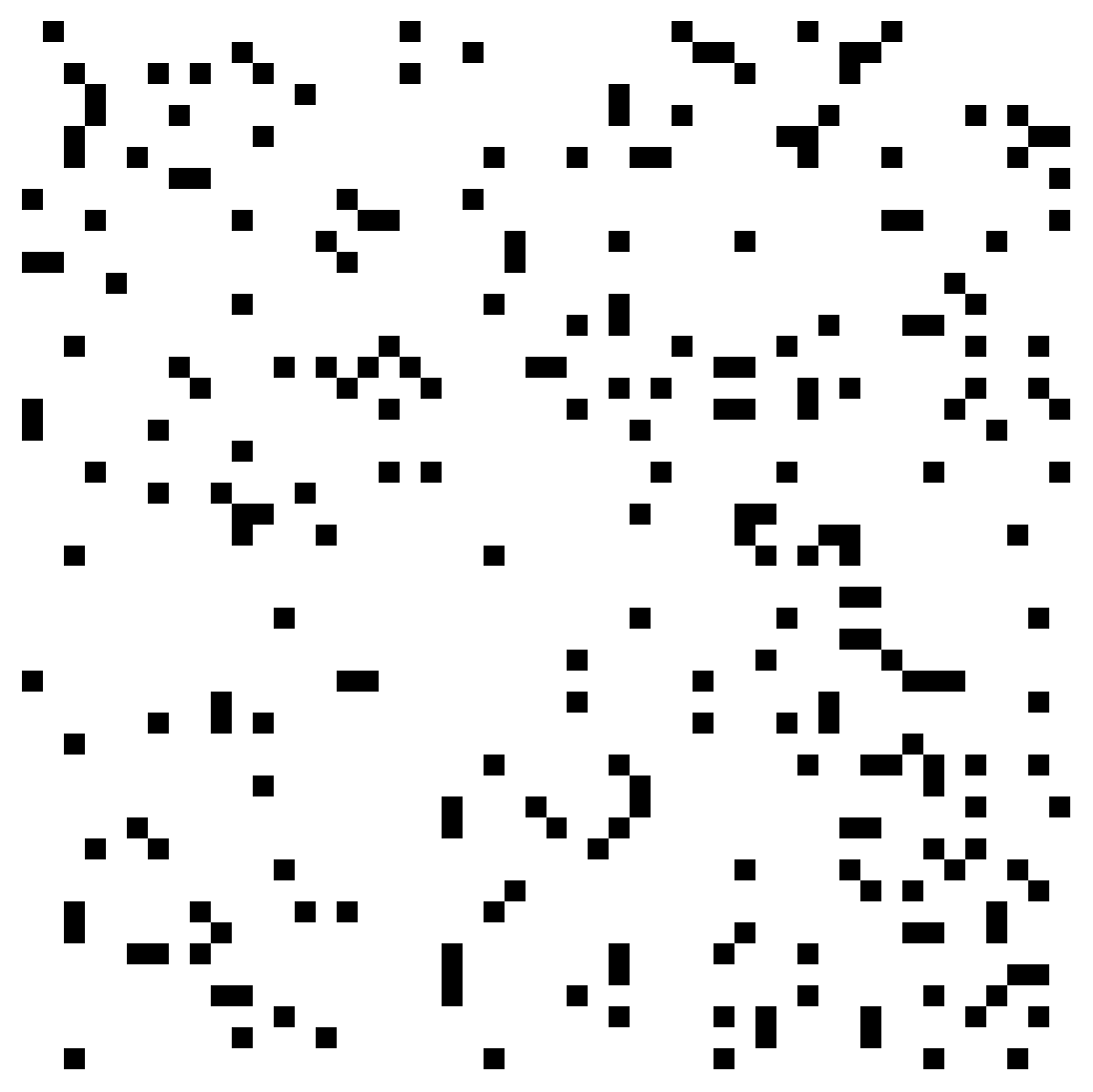}\includegraphics[width=0.25\textwidth]{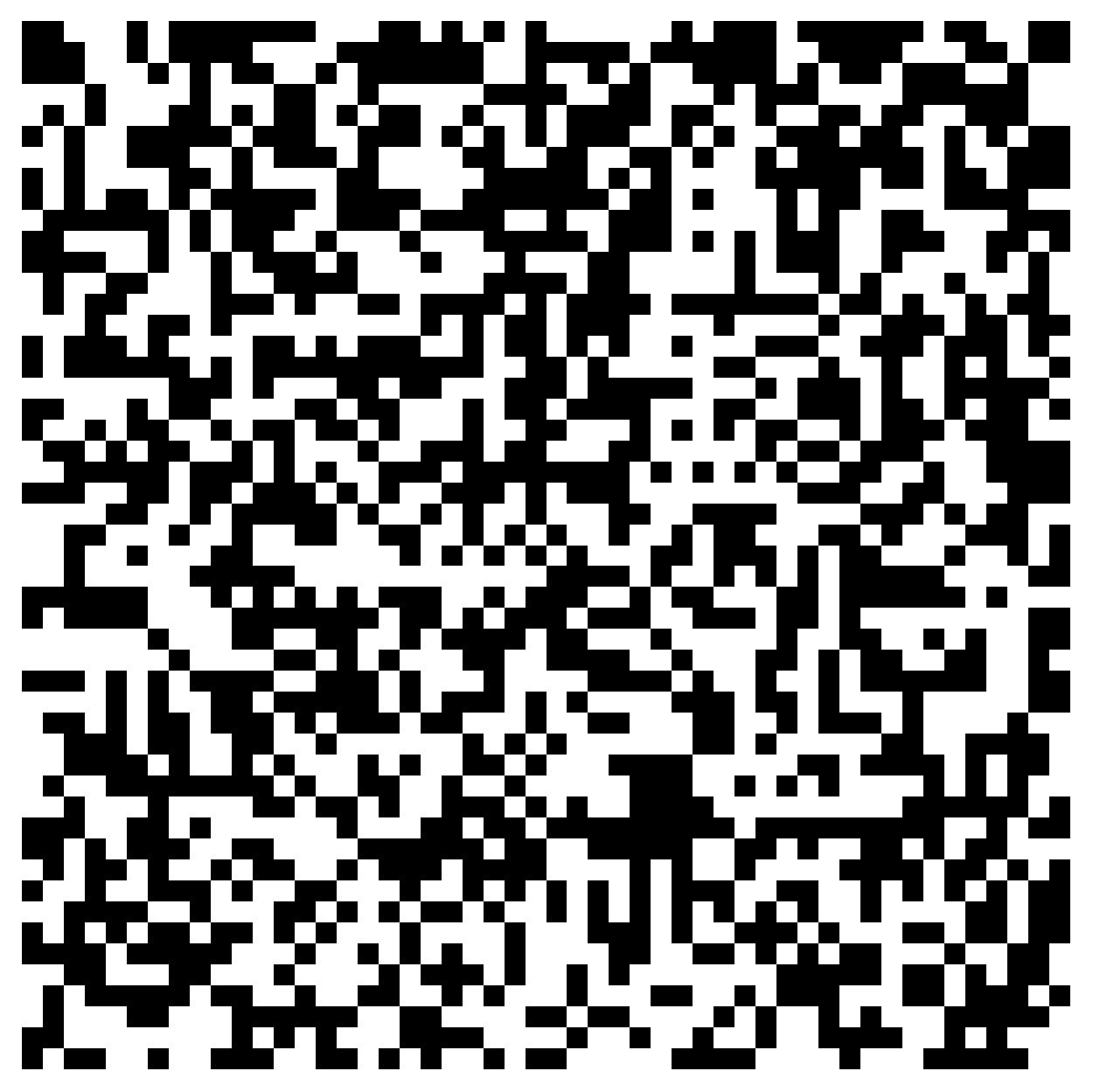}\includegraphics[width=0.25\textwidth]{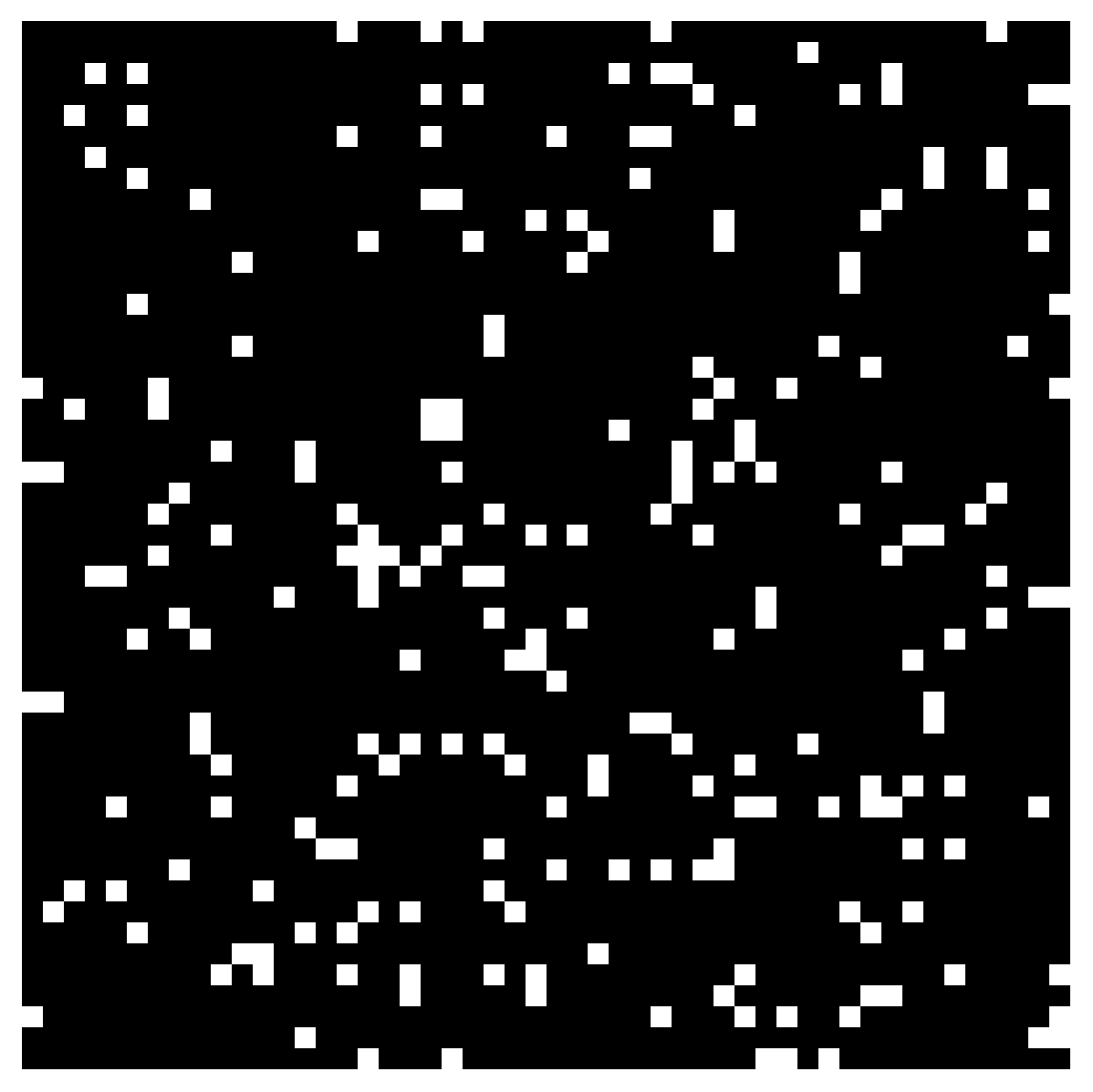}
 \caption{The sublevel filtration for a $50\times 50$ pixel grid. On the left is a pixelized function, with redder pixels corresponding to larger function values. The next three panels depict the sublevel filtration for this function, with the sublevel set in black and the threshold increasing from left to right. Early in the filtration, there are many distinct connected components (contributing to $\beta_0$) and not very many loops ($\beta_1$). As the threshold is increased, formerly distinct sections are connected to each other, and more loops form. For large threshold values, there is only one connected component. There are many loops, which are filled in as the threshold approaches the function's maximum value on the grid.}\label{fig:sub2d}
 \end{figure}
\end{subsection}
\begin{subsection}{Persistence diagrams}\label{PDs}
Persistent homology has a convenient representation in the form of persistence diagrams (PDs). PDs are scatter plots giving the birth and death parameters of homologically distinct $k$-cycles. For our sublevel filtration, there will be two PDs, one for 0-dimensional features (connected components) and one for 1-dimensional features (loops).

For two-manifolds, there are three Minkowski Functionals (MFs) naturally defined in terms of the sublevel parameter $\nu$. The first is the fraction of area of the base manifold $\mathbb{M}$ included in the sublevel set $f^{-1}(-\infty,\nu]$. The second is the total length of the boundaries of the sublevel set. The third, to which PDs directly relate, is called the genus. It is, up to numerical factors, the Euler characteristic $\chi(\nu)$ of the sublevel set. The Euler characteristic is given by the alternating sum of Betti numbers at the same threshold, $\chi(\nu)=\sum_{k\geq0}(-1)^k\beta_k(\nu)$. 

The Euler characteristic of sublevel sets of a random Gaussian field is in fact known analytically \cite{adler2009random}, and thus lends itself well to testing non-Gaussianity in the CMB. However, the transformation from $\{\beta_0(\nu),\beta_1(\nu)\}$ to $\chi(\nu)$ is not invertible, and thus destroys information.  Using this fact, \cite{2012ApJ...755..122C} proposed strengthening the genus curve statistic by instead using the Betti number curves\footnote{For finite grids with finite pixel size, these `curves' are in fact piecewise constant.} themselves. One drawback is that one must then rely on simulations, since even for random Gaussian fields, analytic results are not known for the Betti number curves. However, \cite{2012ApJ...755..122C} was able to demonstrate explicitly that the Betti number curves contain more information about non-Gaussianity than the genus curve for several models.

In this paper, we propose strengthening the Betti number curves as statistics for non-Gaussianity by using information encoded in PDs from the sublevel filtration. It is simple to calculate the Betti number curves using PD data:
\begin{align}
	\beta_k(\nu)=\sum_{b\leq \nu<d} O_k(b,d)
\end{align}
Here $O_k(b,d)$ represents the number of points in the $k$-th PD with birth parameter $b$ and death parameter $d$. We are simply counting distinct cycles that have been `born' and are not yet `dead' at $\nu$. In other words, the Betti number curves are non-invertible linear combinations of PD data. Thus our intuition tells us that the PDs contain strictly more information than the Betti number curves. Specifically, the information in a persistence diagram concerns the tracking of individual cycles. An example of two distinct PDs giving rise to the same Betti number curve is shown in Figure \ref{fig:PDBet}. In this paper we quantify how much more information about non-Gaussianity is encoded in the PDs for a specific set of CMB simulations.

\begin{figure}
\centering
\includegraphics[width=0.5\textwidth]{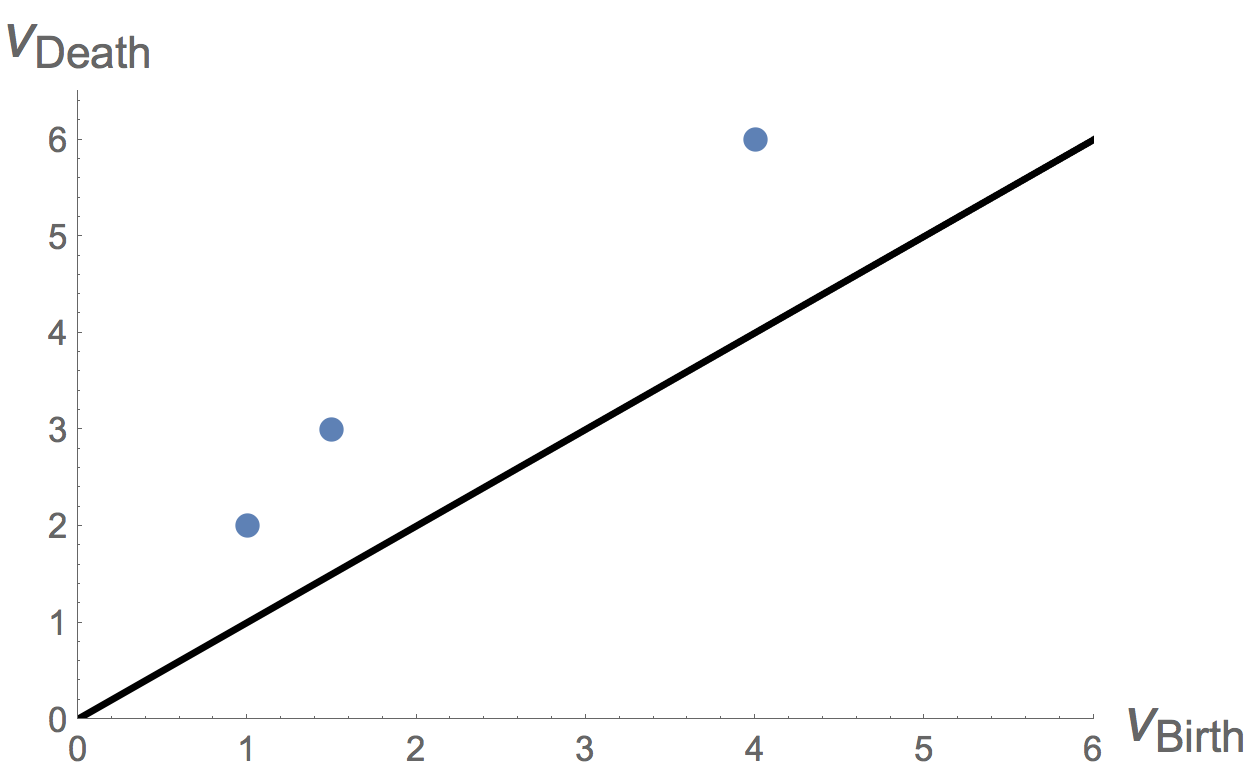}\includegraphics[width=0.5\textwidth]{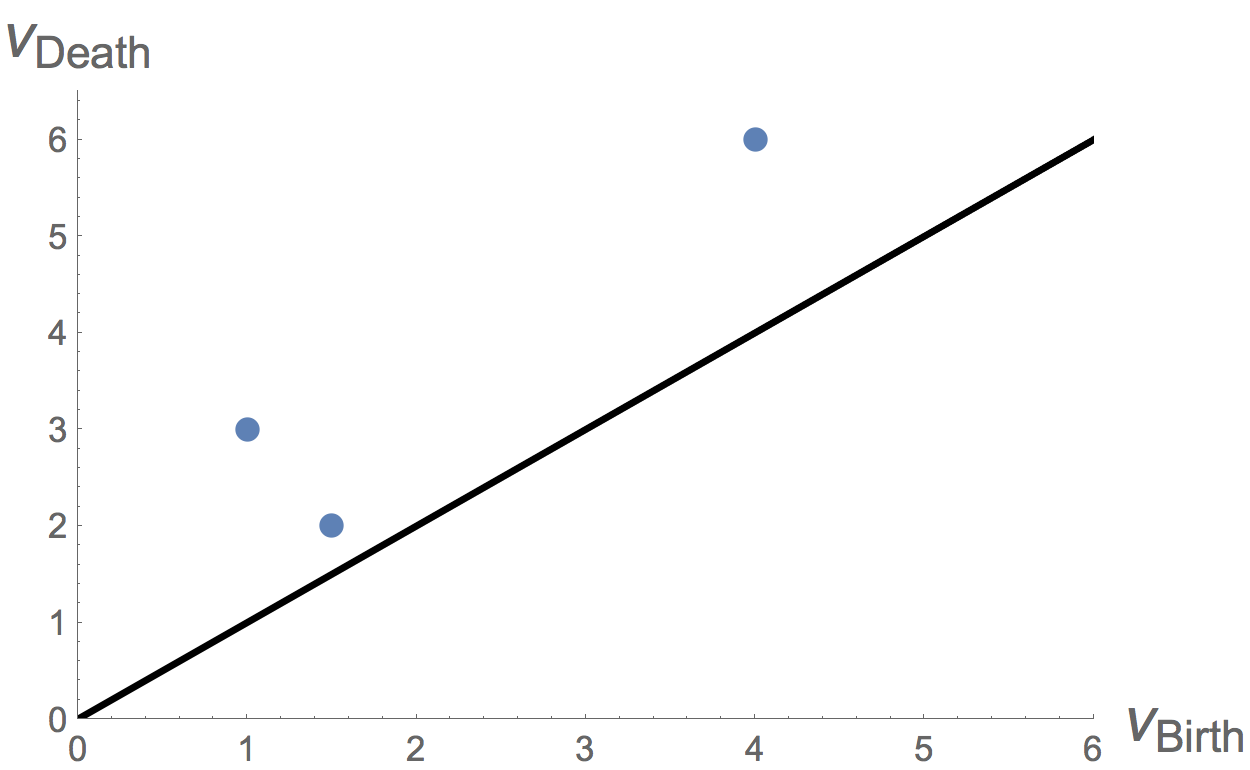}
\includegraphics[width=0.5\textwidth]{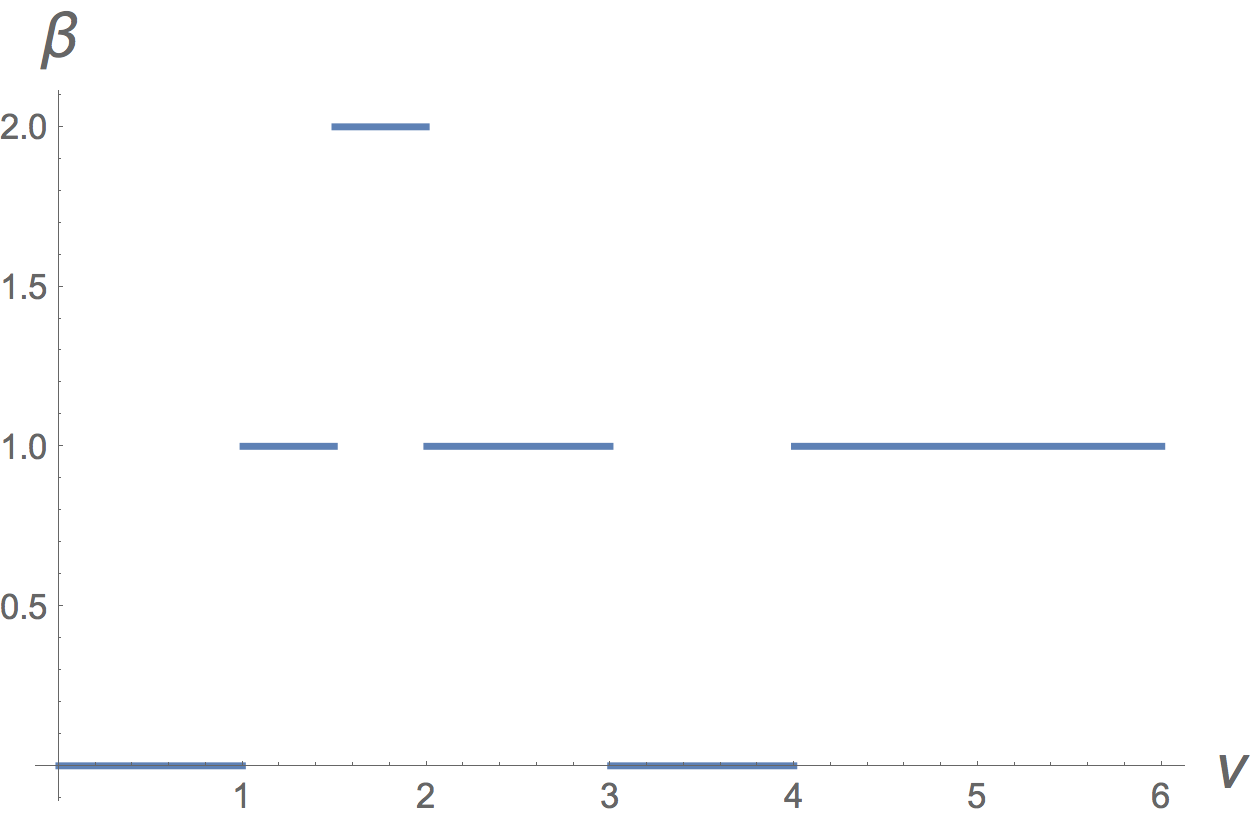}
\caption{The two PDs both give rise to the Betti number curve below. In other words, the Betti number curve can't tell the difference when $\nu_{\rm Death}$ is switched for two cycles, while the PDs are sensitive to this change.}\label{fig:PDBet}
\end{figure}
\end{subsection}
\end{section}
\begin{section}{Local non-Gaussianity}
Local primordial non-Gaussianity \cite{Komatsu:2001rj} takes the ansatz
\begin{align}\label{eqn:ansatz}
	\Phi(\vec{x})=\Phi^G(\vec{x})+f_{NL}\left(\Phi^G(\vec{x})^2-\braket{\Phi^G(\vec{x})^2}\right)
\end{align}
for the primordial gravitational potential $\Phi$. Here $\Phi^G$ is a random Gaussian field and $f_{NL}$ quantifies deviation from Gaussianity. From the primordial gravitational potential, one calculates the CMB temperature anisotropy spherical harmonic expansion $\frac{\Delta T}{T}=\sum_{\ell m} a_{\ell m}Y_{\ell m}$ via
\begin{align}\label{eqn:alm}
	a_{\ell m}=4\pi(-i)^{\ell}\int \frac{d^3 k}{(2\pi)^3}\Phi(\vec{k})\Delta_{\ell}(k)Y_{\ell m}^*(\hat{k})
\end{align}
where $\Delta_\ell(k)$ is the transfer function for temperature in momentum space.
\begin{subsection}{Simulations}\label{sec:sim}
We use the publicly available simulations of local non-Gaussianity provided by Elsner and Wandelt \cite{0067-0049-184-2-264}. Following the method of \cite{2003ApJ...597...57L}, the simulations generate purely Gaussian gravitational potentials. The spherical harmonic coefficients are then calculated by combining the ansatz (\ref{eqn:ansatz}) with (\ref{eqn:alm}).

The resolution of the simulations is determined by $\ell_{\rm max}=1024$, corresponding to HEALPIX \cite{gorski2005healpix} variable $N_{\rm side}=512$. (This is rather low-resolution compared to $N_{\rm side}=2048$ for Planck 2015 data \cite{Ade:2015ava}.) We include a Gaussian smoothing of $\theta_s=90'$, corresponding to full width half maximum FWHM=$\sqrt{8\ln 2}\theta_s=211.934'$. 
The simulations use cosmological parameters from WMAP5+BAO+SN data\footnote{Available at \href{http://lambda.gsfc.nasa.gov/product/map/dr3/parameters.cfm}{http://lambda.gsfc.nasa.gov/product/map/dr3/parameters.cfm}}: $\Omega_\Lambda=0.721,~\Omega_c h^2=0.1143,~\Omega_b h^2=0.02256,~\Delta_{\mathcal{R}}^2(0.002~{\rm Mpc}^{-1})=2.457\cdot 10^{-9},~h=0.701,~n_s=0.96,$ and $\tau=0.084$.

The simulations provide 1000 sets of Gaussian spherical harmonic coefficients $a_{\ell m}^L$ and corresponding non-Gaussian $a_{\ell m}^{NL}$. One can tune the amount of non-Gaussianity in a particular map by using $a_{\ell m}=a_{\ell m}^L+f_{NL}a_{\ell m}^{NL}$. In this paper, we consider simulations with $f_{NL}=0,10,25,50,$ and 100.
\end{subsection}
\end{section}
\begin{section}{Persistence of CMB temperature anisotropy maps}
In this section we describe how persistent homology may be applied to the simulations described in Section \ref{sec:sim}. We outline our numerical pipeline for generating and analyzing sections of CMB temperature anisotropy maps. We then provide numerical results demonstrating that PDs include significantly more information about local primordial non-Gaussianity than the Betti number curves. Using a likelihood function analysis, we show that PDs can constrain $\Delta f_{NL}^{\rm loc}=35.8$ at 68\% confidence on the simulation set.

\begin{subsection}{Persistence and local non-Gaussianity}
For each map, we consider rectangular regions (corresponding to four fundamental pixels in the HEALPIX scheme) centered around the north pole and the south pole. For real CMB data, it makes sense to consider these regions instead of the entire sphere due to the galactic plane. Thus global properties of the sphere should not be taken into account. For each $f_{NL}$ value, we have 2000 grids, each having dimensions $1024\times 1024$ in pixel space.

We use the \texttt{R} package \texttt{TDA} \cite{fasy2014introduction} to perform the sublevel filtration and persistent homology analysis of each grid\footnote{The persistent homology calculation in the package is performed using the Persistent Homology Algorithm Toolkit (\texttt{PHAT}) \cite{bauer2017phat}.}. For each grid, we have two PDs, corresponding to 0-dimensional homology and 1-dimensional homology. Histograms corresponding to the PDs are shown in Figures \ref{fig:PD0} and \ref{fig:PD1} for $f_{NL}=0,100$. For aesthetic reasons, we plot the persistence, defined as $\nu_{\rm death}-\nu_{\rm birth}$, on the vertical axis instead of the death parameter. For the 0-dimensional histograms, there is always at least one connected component. In a true PD, we would represent this feature as a point with infinite persistence. Instead, we artificially assign this point a finite persistence equal to the temperature range of the grid under consideration, $p=T_{\rm max}-T_{\rm min}$. These points give rise to the upper shape in the 0-dimensional histograms.

We also show the differences between histograms with $f_{NL}=50,0$ and $f_{NL}=100,0$ in Figures \ref{diff0} and \ref{diff1} . It is worth convincing oneself that local non-Gaussianity affects PDs in a systematic way, i.e. that these diagrams display some pattern rather than just random noise. Since we are in the regime of weak non-Gaussianity, we can assume that turning on $f_{NL}$ does not create or destroy critical points, so the total number of homological features is approximately unchanged. Instead, we should look for how features move in the PDs as $f_{NL}$ is increased. In these histograms, blue bins lose features and red bins gain features as $f_{NL}$ is increased. We see that the effect of local non-Gaussianity on the 0-dimensional histogram is to move features down and to the left, with earlier births and shorter persistences. For the 1-dimensional histogram we see the features move to shorter persistences, with less of an effect on birth times.

We also calculate the Betti number curves and the effect of local non-Gaussianity on them, shown in Figure \ref{betti}. We observe the same pattern for $\beta_i(f_{NL}=100)-\beta_i(f_{NL}=0)$ found in \cite{2012ApJ...755..122C}.
\begin{figure}[h!]\centering
\includegraphics[width=0.50\textwidth]{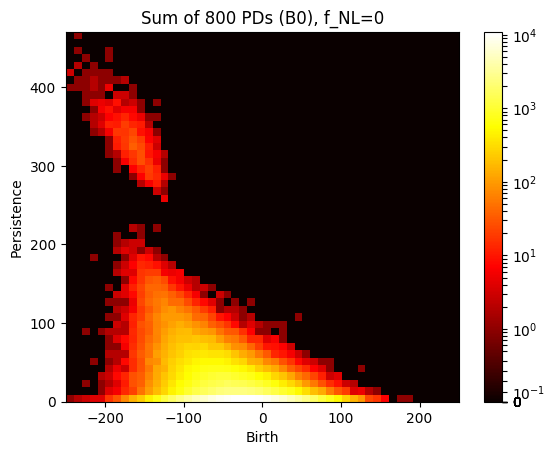}\includegraphics[width=0.50\textwidth]{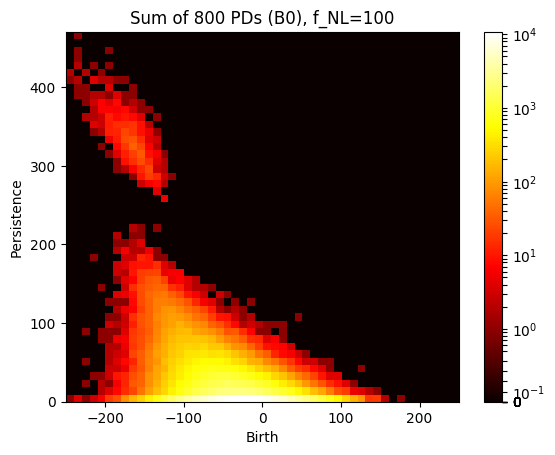}
\caption{Persistence diagrams for 0-dimensional features, binned as 2-dimensional histograms. We artifically assign a finite persistence $p=T_{\rm max}-T_{\rm min}$ to the earliest-born component on the grid, resulting in two distinct shapes. The effect of $f_{NL}=100$ is to give $0$-cycles earlier births and shorter persistences.}\label{fig:PD0}
\end{figure}

\begin{figure}[h!]\centering
\includegraphics[width=0.50\textwidth]{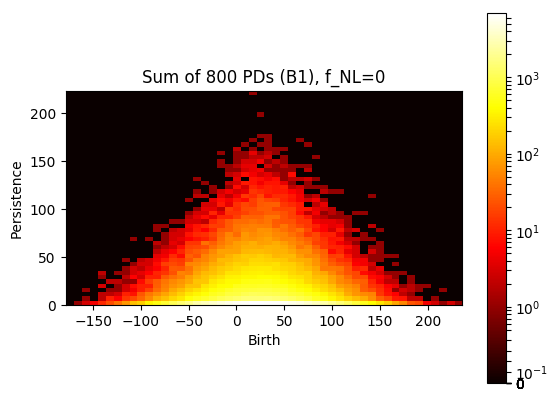}\includegraphics[width=0.50\textwidth]{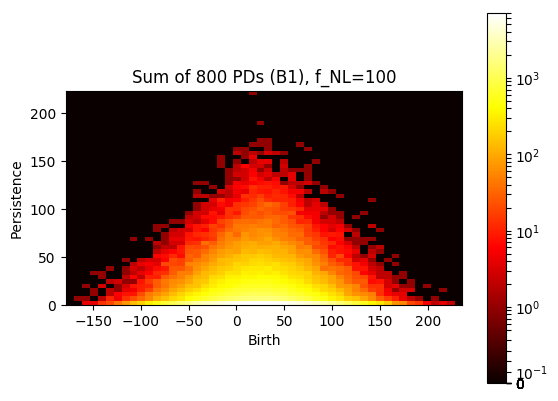}
\caption{Persistence diagrams for 1-dimensional features. The effect of $f_{NL}$ is to give 1-cycles shorter persistences.}\label{fig:PD1}
\end{figure}

\begin{figure}[h!]\centering
\includegraphics[width=0.50\textwidth]{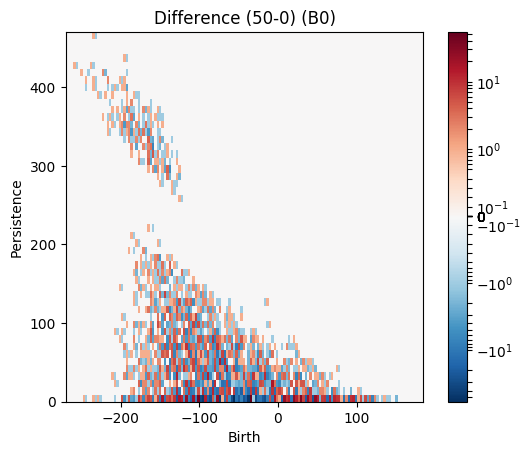}\includegraphics[width=0.50\textwidth]{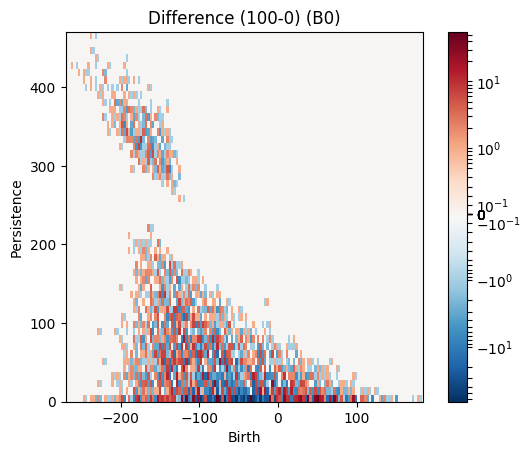}
\caption{Histograms showing the movement of 0-cycles under the influence of local non-Gaussianity, $PD_0(f_{NL}=50)-PD_0(f_{NL}=0)$ (left) and $PD_0(f_{NL}=100)-PD_0(f_{NL}=0)$ (right). When $f_{NL}$ is raised, cycles leave the blue bins and enter the red bins. On the whole, we see that 0-cycles tend to move to earlier births and shorter persistences. We also observe a fairly large blue region at short persistence and small negative birth parameter.}
\label{diff0}
\end{figure}

\begin{figure}[h!]\centering
\includegraphics[width=0.50\textwidth]{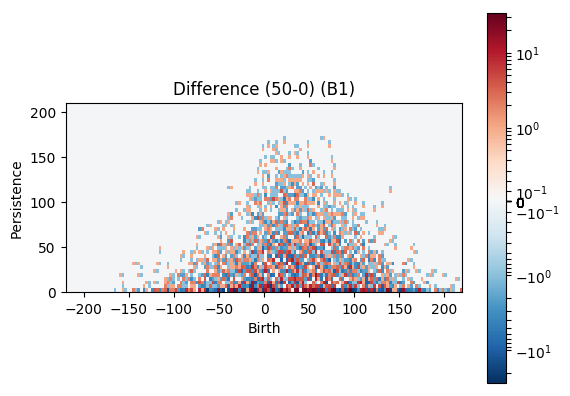}\includegraphics[width=0.50\textwidth]{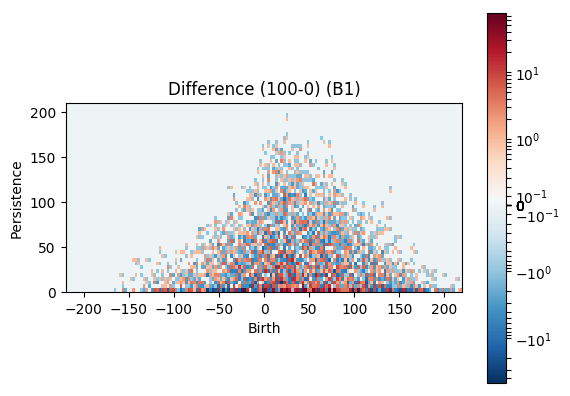}
\caption{Histograms showing the movement of 1-cycles under the influence of local non-Gaussianity, $PD_1(f_{NL}=50)-PD_1(f_{NL}=0)$ (left) and $PD_1(f_{NL}=100)-PD_1(f_{NL}=0)$ (right). When $f_{NL}$ is raised, cycles leave the blue bins and enter the red bins. Generally, cycles move to shorter persistences.}
\label{diff1}
\end{figure}

\begin{figure}[h!]\centering
\includegraphics[width=0.45\textwidth]{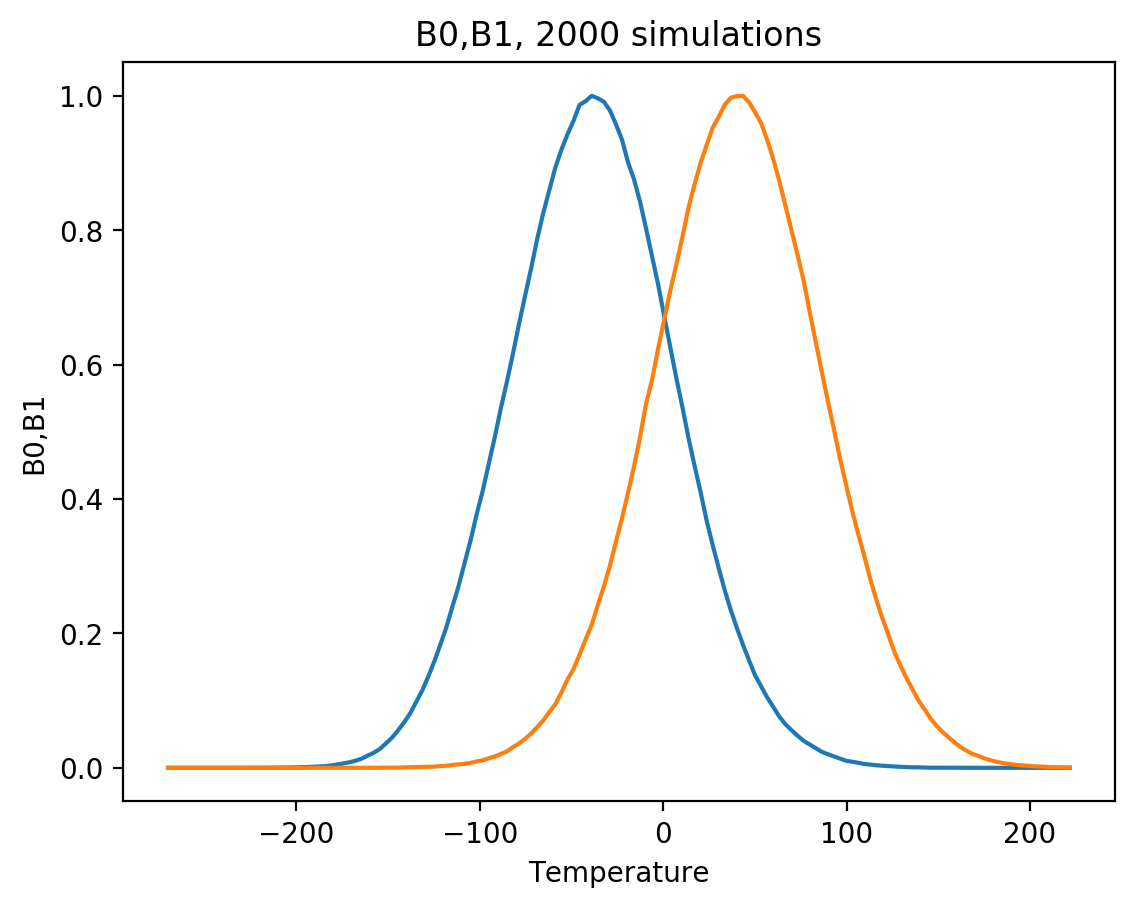}
\includegraphics[width=0.45\textwidth]{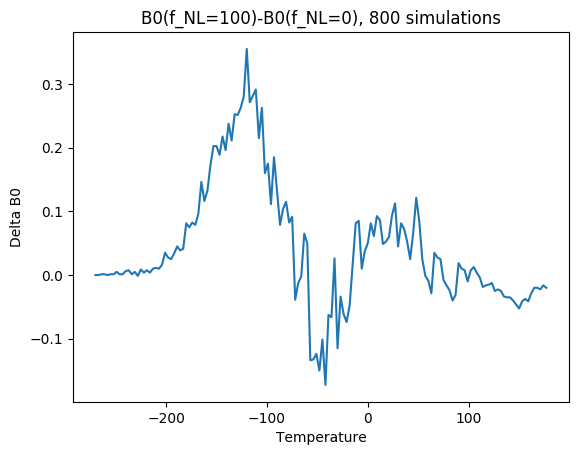}\includegraphics[width=0.45\textwidth]{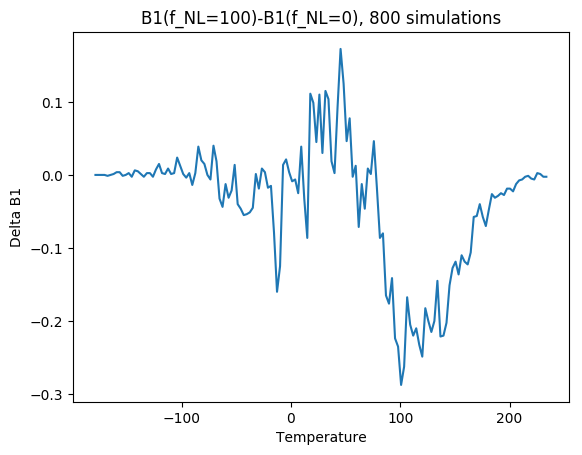}
\caption{Betti number curves for our simulations. On top, $\beta_0$ (blue) and $\beta_1$ (orange) as functions of temperature. Here we normalize by $\beta_{\rm max}$ since $\beta_0$ is artificially larger due to the boundary of the rectangular patch. Below we have $\Delta\beta_i=\left(\beta_i(f_{NL}=100)-\beta_i(f_{NL}=0)\right)/N_{\rm sim}$. The difference plots have the same characteristic shapes observed in \cite{2012ApJ...755..122C}.}\label{betti}
\end{figure}
\end{subsection}
\begin{subsection}{Statistical content of persistence diagrams}
The likelihood function $P({\bf d}|{\bf p})$ is a standard tool in cosmological data analysis, used to determine best-fit parameter values and errors. Here $P$ gives the probability that a set of parameters ${\bf p}$ would give rise to a data vector ${\bf d}$. Via Bayes' theorem, one has
\begin{align}
	P({\bf p}|{\bf d})=\frac{P({\bf d}|{\bf p})P({\bf p})}{P({\bf d})}
\end{align}
The denominator is a simple normalization factor, and thus does not affect the location of the likelihood function's peak or its width. Assuming a uniform prior $P({\bf p})$, one then has that $ P({\bf p}|{\bf d})\propto P({\bf d}|{\bf p})$. (This is a useful proportionality, since the goal of data analysis is to constrain ${\bf p}$ by measuring ${\bf d}$.) We may get some sense for the information content of our PDs by assuming a Gaussian likelihood function
\begin{align}
	P({\bf d}|f_{NL})=\frac{1}{2\pi\sqrt{\det {\bf C}}}\exp\left(-\frac{1}{2}({\bf d}-\boldsymbol{\mu}){\bf C}^{-1}({\bf d}-\boldsymbol{\mu})\right)
\end{align}
where ${\bf d}$ is our data vector, $\boldsymbol{\mu}=\boldsymbol{\mu}(f_{NL})$ is the model data vector at a given $f_{NL}$, and ${\bf C}$ is the covariance matrix. This is a good approximation when the fluctuations of each component of the data vector are small compared to their ensemble average. To stay in this regime, we must throw out some of our bins. This corresponds to ignoring sufficiently rare events. For Betti number curves, the data vector's entries are the curve's values sampled at certain temperatures, while for the PDs the data vector's entries are the number of points in particular bins of the corresponding 2-dimensional histogram. The sharper the likelihood function is peaked at its maximum, the more sensitive it is to the cosmological parameters under consideration. This intuition is formalized in the Fisher matrix determination of error \cite{Tegmark:1996bz}. Along these lines, we suppose our data comes from a theory with $f_{NL}=0$, using the average data vector over 800 simulations. We then evaluate the likelihood function for each statistic using $f_{NL}=0,10,50$ for $\boldsymbol{\mu}(f_{NL})$, also averaging over 800 simulations. Here we calculate the covariance matrix ${\bf C}$ using 2000 $f_{NL}=0$ simulations. (We also include a correction factor related to sample size for the inverse covariance estimator \cite{Hartlap:2006kj}.) To a good approximation, the likelihood function is Gaussian in $f_{NL}$. We thus fit the resulting likelihood function with a Gaussian to determine our 68\% confidence constraint. The results are shown in Table 1. We find that combining $PD_0$ and $PD_1$ gives a $\Delta f_{NL}=35.8$ constraint at 68\% confidence on the simulation set, which seems reasonable compared to a detailed forecasting (albeit for higher-resolution maps) for the MFs like \cite{2013MNRAS.429.2104D}. We observe that the PDs give constraints almost twice as strong as the Betti number curves, which are in turn \cite{2012ApJ...755..122C} stronger than the genus.

\begin{center}
  \begin{tabular}{ l | r }
    \hline
    Statistic & $\Delta f_{NL}$ \\ \hline
    $\beta_0$ & 67.4  \\ \hline
    $\beta_1$ & 66.1 \\ \hline
    $\beta_0+\beta_1$&60.6\\ \hline
    $PD_0$ & 39.1 \\ \hline
    $PD_1$ & 37.4 \\\hline
    $PD_0+PD_1$ & 35.8\\\hline
  \end{tabular}

  Table 1: Relative information content of persistence diagrams.
\end{center}

\begin{figure}
\centering
\includegraphics[width=0.5\textwidth]{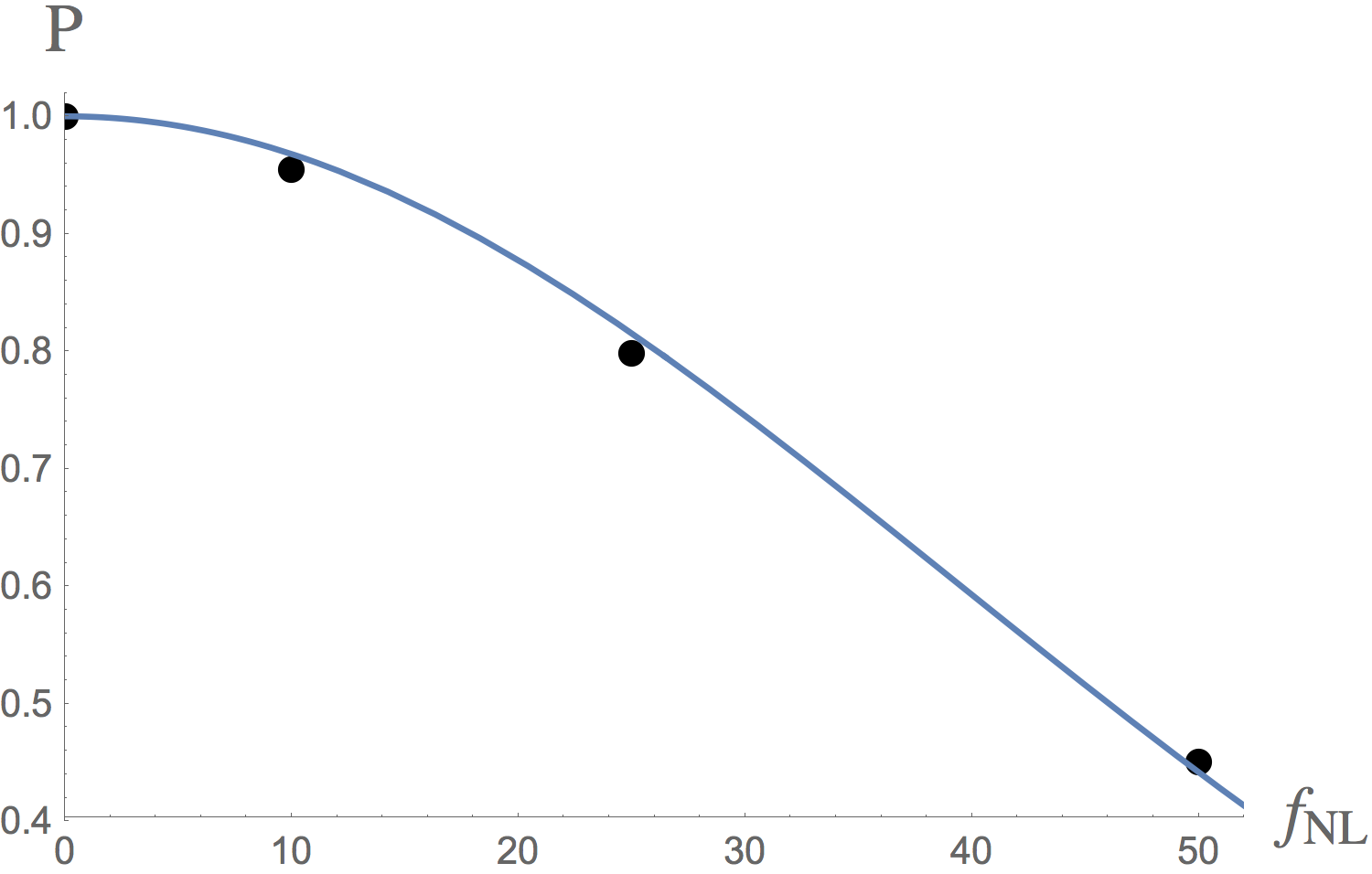}\includegraphics[width=0.5\textwidth]{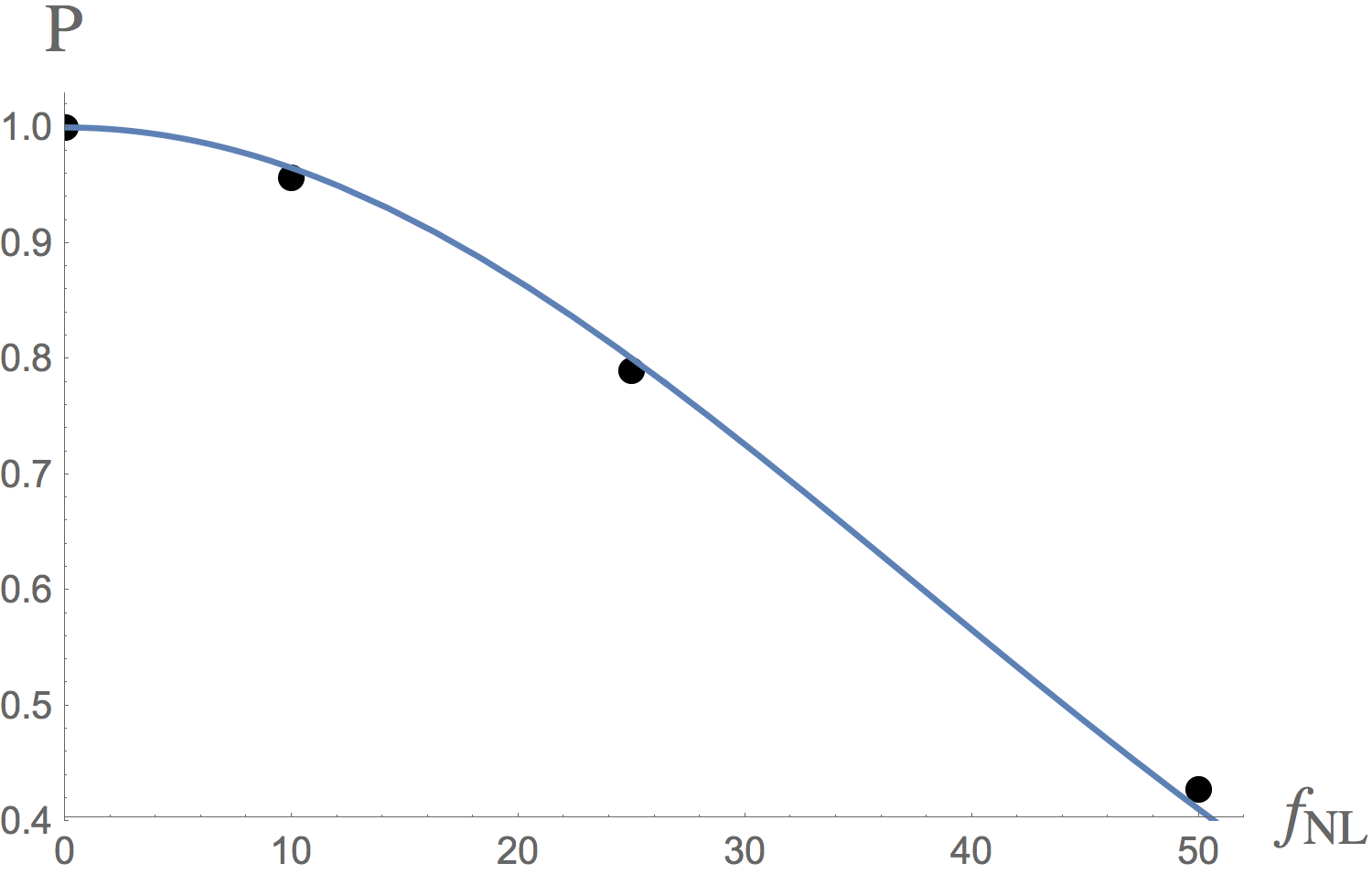}
\caption{Gaussian fits of the likelihood functions for $PD_0$ (left) and $PD_1$ (right), giving $\sigma=39.1,37.4$, respectively.}
\end{figure}

\end{subsection}
\end{section}
\begin{section}{Conclusions and Outlook}
In this paper, we have introduced the persistence diagram (PD) as a useful statistical observable for CMB temperature anisotropy data. We propose that, in addition to constraining other cosmological parameters, PDs may be useful in the search for non-Gaussianity. We have computed the PDs for a set of simulated CMB temperature maps with local non-Gaussianity, using the sublevel filtration. We found that the information contained in the PDs could discriminate between different levels of local non-Gaussianity more sensitively than the Betti number curves. 
This is a sensible result, as the Betti number curves can be derived from the PDs, but not vice versa. In particular, we expect PDs to be able to constrain local non-Gaussianity to $\Delta f_{NL}=35.8$ at 68\% confidence.

Our computation being a proof of concept, we have ignored several effects. We did not take into account observational effects like sky cuts, instrument noise, or beam width. We chose a smoothing angle of $\theta_s=90'$, and did not study the effects of varying this. We also focused on the temperature anisotropy, while in principle there is more information contained in the topology of CMB polarization \cite{Ganesan:2014lqa,Chingangbam:2017sap}. We also only considered non-Gaussianity of local type, and did not simultaneously vary other cosmological parameters. It would be particularly interesting to see how PDs compare with the bispectrum for shapes of non-Gaussianity that are particularly difficult to constrain using templates, like the oscillatory shapes arising from axion monodromy \cite{Flauger:2014ana}. As the present work aims to provide a proof of concept, we defer these interesting issues to our future work.

Our results indicate that persistent homology could also be useful in other situations where topological observables have been used. In particular peak counts (Betti numbers curves in disguise) have been proposed to derive constraints on cosmological parameters using CMB lensing data \cite{Liu:2016nfs}. The methods presented in this paper could also be useful in studying the topology of large scale structure. In this case, persistent homology has been proposed to identify various structures in the cosmic web \cite{Pranav:2016gwr,2011MNRAS.414..350S,2011MNRAS.414..384S}. It would be interesting to compare a method using persistent homology to a topologically-flavored algorithm like ZOBOV \cite{Neyrinck:2007gy} for the purposes of void identification. It is interesting to note that similar clustering and void features also arise in the distribution of string vacua \cite{Ashok:2003gk,Denef:2004ze,Giryavets:2004zr,DeWolfe:2004ns}. 
However, when the dimension of the moduli space (the data space in this case) is huge, we can no longer visualize the distribution of vacua. Nonetheless, we can compute the persistent homology of point clouds generated by string vacua and thus diagnose the shape of this data.
(Some preliminary work along these lines can be found in  \cite{Cirafici:2015pky}.) We plan to report our findings in this interesting direction in a forthcoming work \cite{WIP}. 


It is worth noting the relation of this paper to \cite{2017arXiv170408248A}, which aims to put TDA on a statistically sound footing. Their method circumvents the need for simulations by using just one PD, then generating statistically similar PDs. They then perform a toy example with Planck CMB data, treating the north and south caps separately. The north and south caps give different results in their framework, leading them to claim that their data should not be regarded as arising from the same stochastic process. This paper has instead treated PDs as statistics in themselves, and demonstrated how topological methods are strengthened in this approach.

\end{section}

\subsection*{Acknowledgments}
We would like to thank Amol Upadhye and Jon Brown for helpful discussions.
This work is supported in part by the DOE grant DE-SC0017647
and the Kellett Award of the University of Wisconsin.

\bibliography{CMB_Persistence.bib}\bibliographystyle{utphys}

\end{document}